# Toward AI-Agent-Driven Particle Transport Simulations: Implementation of AI-Assisted Workflows for PHITS

*Tatsuhiko Sato[a*], Shintaro Hashimoto[a], Tatsuhiko Ogawa[a], Takuya Furuta[a], Yuho Hirata[a], Seiki Ohnishi[a], Yasuhito Sakaki[b], Nobuhiro Shigyo[c]*

[a]*Japan Atomic Energy Agency, Shirakata 2-4, Tokai, Ibaraki 319-1195, Japan;* [b]*High Energy Accelerator Research Organization, Oho 1-1, Tsukuba, Ibaraki, 305-0801, Japan;* [c]*Kyushu University, Department of Applied Quantum Physics and Nuclear Engineering, Motooka 744, Nishi, Fukuoka, 819-0395, Japan;*

*Corresponding author. Email: sato.tatsuhiko@jaea.go.jp

## Toward AI-Agent-Driven Particle Transport Simulations: Implementation of AI-Assisted Workflows for PHITS

Monte Carlo particle transport codes are powerful tools, but their use requires substantial knowledge of input preparation, execution, and result analysis. In this study, we present a code-side strategy for applying existing AI assistants and AI agents to PHITS. Two complementary sets of AI-ready resources were prepared from manuals, lecture materials, sample inputs, utility information, and developer-curated cautions: a bundled knowledge base for retrieval-augmented generation (RAG)-based assistants and a compact agent reference for direct use by AI agents. The knowledge base was loaded into Gemini Notebook to provide conversational PHITS support, while the agent reference was combined with PHITS-specific policies and execution rules to enable AI agents to edit input files, execute calculations, inspect errors, analyze results, and assist with source-code modification and compilation. To evaluate the agent-driven workflow, five demonstration tasks were executed ten times with different AI agents and computing environments. The results showed that AI agents could handle complex PHITS workflows when appropriate resources and rules were provided. Practical lessons included precise prompts, human verification, well-documented sample files, explicit execution policies, and command-line-accessible tools. These findings support bundling AI-ready resources with particle transport codes to enable the use of general-purpose AI tools without requiring dedicated code-specific applications.

### 1. Introduction

Monte Carlo particle transport codes are widely used in accelerator facility design, nuclear technology, medical physics, space radiation dosimetry, and many other fields. General-purpose codes such as FLUKA [1], Geant4 [2], MCNP [3], OpenMC [4], PHITS [5], and TRIPOLI-4 [6] are powerful tools for detailed simulations of the transport of neutrons, photons, electrons, protons, heavy ions, and other particles in complex geometries. However, appropriate use of these codes requires substantial knowledge and experience, including input syntax, geometry modeling, material definition, source specification, tally selection, statistical uncertainty evaluation, and output analysis.

Recent advances in large language models (LLMs) have rapidly expanded the use of natural-language interfaces for program generation, file editing, error correction, and data analysis [7]. In addition, LLM-based autonomous agents, often referred to as AI agents, have attracted increasing attention as a means of combining reasoning, planning, tool use, and iterative task execution [8]. Retrieval-augmented generation (RAG) further enables LLMs to answer questions while referring to external documents rather than relying solely on their internal parameters [9]. These technologies may also be useful for assisting users of particle transport codes.

In the field of radiation-transport Monte Carlo simulations, AutoFLUKA has recently been reported as an LLM-agent framework for automating FLUKA workflows, including input preparation, simulation execution, and post-processing [10]. It is an important precedent showing that AI agents can be applied to Monte Carlo simulation workflows, and its authors suggested that such frameworks could be extended to other simulation packages. For such extensions to be realized in practice, however, code developers need to prepare and distribute AI-ready code-side resources, such as knowledge bases, sample-file descriptions, reference policies, and execution rules, that allow general-purpose AI assistants and AI coding agents to use each code reliably.

Motivated by this perspective, we present an implementation strategy for applying existing LLMs and AI agents to particle transport simulations through the development of AI-assisted PHITS workflows. To achieve this, the following tasks were carried out: (1) PHITS manuals, lecture materials, sample input files, and other instructive materials were reorganized into AI-ready text resources; (2) a conversational support environment was constructed by loading these text resources into Gemini Notebook, a RAG-based AI assistant; and (3) an autonomous PHITS execution and analysis workflow was developed using AI agents such as Codex and Claude Code by making the text resources and additional PHITS-specific operational rules available to them. It should be noted that the primary purpose of this study is not to develop a new LLM algorithm, nor a dedicated AI application for PHITS, but to establish a practical methodology for preparing the code-side AI-ready resources and operational rules that allow general-purpose AI assistants and AI agents to operate particle transport simulation codes.

## 2. Methods

### *2.1. Overview of the PHITS package with AI-assisted workflow components*

Figure 1 shows an overview of the PHITS package with AI-assisted workflow components introduced in this study. The PHITS package contains several directories that support users in preparing input files, such as *lecture*, *manual*, *recommendation*, *sample*, and *utility*. These directories provide lecture materials, PHITS and supporting tool manuals in HTML format, recommended input files for typical simulations, sample input files and user-defined programs, and supporting tools such as PHITS Interactive Geometry Viewer in 3D (PHIG-3D) [11] and the RadioTherapy package based on PHITS (RT-PHITS) [12,13]. The package also contains other directories for executables, source code, and cross-section libraries, although they are not shown in the figure for simplicity.

This study focuses on the *workbench* directory, an AI-oriented directory newly introduced in PHITS version 3.37, which was released on September 1, 2026. The *workbench* directory contains AI-ready text resources generated from the PHITS manual and other user-support materials. These resources are intended for use by AI assistants and AI agents when answering PHITS-related questions or searching for practical examples. The directory also provides practical instructions and tools for an AI-agent-driven workflow, including agent entry-point files, reference policies, task definitions, PHITS-input language-support settings, and setup scripts. Through this workflow, AI agents operating in an integrated development environment (IDE) such as VS Code can edit PHITS input files, execute PHITS, analyze output files, and iteratively repeat these procedures until the user's request is satisfied. The preparation and use of the *workbench* directory are described in the following subsections.

*2.2. Preparation of AI-ready PHITS knowledge resources*

To use particle transport codes with AI, it is not sufficient simply to provide AI systems with manuals in HTML or PDF format. This is because manuals primarily serve as comprehensive reference documents; practical knowledge resources with contextual information, such as examples, tutorials, and recommended workflows, are also required. Therefore, we developed a Python script to convert information related to practical PHITS usage contained in the *lecture*, *manual*, *recommendation*, *sample*, and *utility* directories into text formats that can be easily searched and referenced by AI systems. The resulting resources were stored in the *knowledge_base* and *agent_reference* directories under *workbench/AI* for use by RAG-based AI assistants and AI agents, respectively.

In constructing these resources, we adopted the following seven measures to make it easier for AI systems to search, interpret, and use the information:

(a) Overview descriptions were prepared for each source directory to summarize the types of information contained in the converted text files and to help AI systems determine where to search according to the user's calculation purpose.

(b) Keywords were assigned to individual input files and related materials so that AI systems could more easily identify relevant information, such as calculation purposes, particle types, geometries, tallies, and output quantities.

(c) The lecture materials in the *lecture* directory were linked with slide texts, narration scripts where available, and exercise input files so that practical workflows could be referred to together with the corresponding examples.

(d) Information on sample input files and user-defined programs in the *sample* directory was organized together with the relevant instruction texts, where available, allowing AI systems to refer to advanced functions when necessary.

(e) Lecture materials, sample files, and manuals for supporting tools in the *utility* directory were systematically converted into text-based resources so that AI systems could understand how these tools should be used and could refer to their usage procedures.

(f) Developer-curated cautions were summarized as crucial notices to explicitly describe practical points that AI systems may easily misunderstand, such as input syntax, geometry checking, tally selection, and execution procedures.

(g) For direct use by AI agents, the *agent_reference* directory was organized into small, categorized Markdown files. Because AI agents can directly access the original project files when necessary, input files, output files, and other plain-text source files were generally referenced by path rather than duplicated in this directory. This structure enables targeted searches and reduces unnecessary token consumption.

It should be noted that these measures were developed through iterative consultation with AI systems, which provided suggestions on how information should be structured to match their own search and interpretation processes. More detailed strategies for preparing these AI-ready resources are provided in Appendix I.

### *2.3. Conversational PHITS support using a RAG-based AI assistant*

The constructed knowledge base was first loaded into Gemini Notebook, a RAG-based AI assistant, to establish conversational PHITS support in which users can ask questions in natural language. In this approach, users do not need to search manuals, lecture materials, and sample files individually. Instead, they can ask the AI assistant questions such as "Which recommended input file is appropriate for accelerator shielding design?", "What does this keyword mean?", or "What is this sample input intended to do?" This is particularly useful when users are not familiar with the PHITS directory structure.

An advantage of the conversational AI-assistant approach is its ease of introduction. Users can access this environment simply by visiting an existing public website configured with the above knowledge base, named AI Phi-chan [14], and signing in with a Google Account, without preparing a special programming environment. Gemini Notebook also supports multilingual output, allowing users to ask questions and receive explanations in a supported language even when the underlying PHITS resources are available only in English or Japanese. This capability may substantially reduce language barriers for users who are not proficient in either language. This approach is therefore useful for beginner education, self-study after PHITS tutorials, preliminary consultation before preparing input files, and efficient searching of the PHITS manual. Because the knowledge-base files are distributed with PHITS, users can also construct similar dedicated PHITS support assistants on other RAG-based

AI platforms by uploading these files together with additional reference materials, such as lecture materials in their preferred language, project-specific documents, or frequently used input files. This flexibility is particularly useful in organizations or regions where access to Google services is restricted or unavailable, as Gemini Notebook may not be accessible to all users.

However, this conversational support approach also has limitations. In environments such as Gemini Notebook, the main functions are limited to dialogue with users, document summarization, explanation of sample files, and suggestions for input-preparation strategies. Such environments do not directly edit PHITS input files, execute PHITS, correct errors, analyze output files, or visualize calculation results. Therefore, moving from conversational support to automated PHITS calculations requires an agentic workflow that allows AI systems to directly edit files, run commands, and inspect outputs.

### *2.4. AI-agent-driven PHITS execution and analysis workflow*

To develop such an automated PHITS workflow, we prepared the workbench structure so that AI agents could access the text resources, operational rules, task definitions, and user-specific cautions. In this study, Codex, Claude Code, and GitHub Copilot were used as AI agents, and VS Code served as the IDE providing the workspace interface. To use all functions of this workflow, users must install PHITS, a compatible Fortran compiler, an IDE, and the command-line interface (CLI) for at least one supported AI agent. They must also obtain an account or API access for the selected AI service. Depending on the service and usage plan, a paid subscription or usage-based fees may be required; such external AI services are not included in the PHITS package.

After these prerequisites have been satisfied on a Windows, macOS, or Linux computer, the user first executes the setup script in the *workbench/IDE* directory. This script performs the following procedures: (1) it checks the availability of required tools, such as Fortran compilers, and configures VS Code tasks for building and executing PHITS; (2) it installs language-support functions for PHITS input files, such as syntax highlighting, auto-completion, and hover information; (3) it copies the common agent instruction file *workbench/AI/AGENTS.md* to the root of the workspace, i.e., the *phits* directory, as *AGENTS.md* for Codex and duplicates the same file as *CLAUDE.md* for Claude Code, so that both AI agents can follow the same PHITS-specific reference policies, cautions, and working rules; and (4) it creates a template for user-specific crucial notices, named *crucial_notice.txt*, in the user directory.

The common agent instruction file, *AGENTS.md*, was intentionally kept simple. This file instructs the AI agents to read *workbench/AI/reference_policy.md* as the highest-priority document for PHITS-related tasks, to perform targeted searches of *workbench/AI/agent_reference* when PHITS-specific information is required, and to refer to user/crucial_notice.txt for practical cautions. The *reference_policy.md* file provides more detailed procedures for AI-agent-based editing and execution of PHITS calculations. For example, after modifying geometry, the agent is instructed to run PHITS in geometry-check mode and confirm that there are no overlapping or undefined regions before proceeding to an actual particle transport calculation.

When an AI agent is asked to modify a PHITS input file, it first reads the target input file and identifies the relevant sections, keywords, particle names, material numbers, cell numbers, tally names, and other related items. If the user's request is sufficiently clear, the agent modifies only the relevant parts with the minimum necessary changes. If the request is ambiguous, the agent performs targeted searches of *workbench/AI/agent_reference* and other relevant resources, identifies the necessary input sections and keywords, and determines how the input file should be modified.

This procedure was designed to prevent the agent from unnecessarily rewriting unrelated parts of the input file and to help maintain the consistency of PHITS input files.

The AI-agent-driven workflow can also handle more advanced operations. When custom functions, such as user-defined tallies or source routines, are required, the agent prepares the corresponding source files, builds PHITS with them, and runs the resulting executable. When optimization of a calculation parameter is requested, the agent can iteratively modify the relevant conditions, run PHITS, compare the obtained results, and identify the optimal setting. In addition, when an AI agent makes a mistake during input preparation, the corresponding caution can be added to *user/crucial_notice.txt*, either manually by the user or by the agent under the user's instruction. This allows the user to incrementally customize the workflow based on their own experience.

*2.5. Demonstration tasks*

To evaluate the capability of the AI-agent-driven workflow, the following prompts were applied as demonstration tasks.

- ✓ Task 1: Please calculate the response function of a Si detector to cosmic-ray irradiation, referring to *recommendation/CosmicRay*. Set up the following geometry and irradiation conditions: (1) place a 5 cm × 5 cm × 0.1 cm silicon detector inside an aluminum spherical shell, (2) the aluminum shell has an inner radius of 10 cm and an outer radius of 15 cm, (3) irradiate the shell isotropically from the outside with galactic cosmic-ray in DeepSpace mode, (4) consider protons only as the source. Estimate the detector response rate in counts/sec, assuming that the threshold deposition energy in the detector is 1 MeV. The statistical uncertainty of the estimated response rate should be less than 10%.

- ✓ Task 2: Please calculate the optimum polyethylene shielding thickness for a neutron source, referring to *recommendation/h10multiplier*. Set up a 10 MBq Cf-252 source at the center and surround it with a spherical polyethylene shield. Variance reduction is not required. Calculate the ambient dose equivalent rate, $H^*(10)$, on the outer surface of the polyethylene sphere. Find the radius of the polyethylene sphere required to reduce $H^*(10)$ to 5 μSv/h or less. Adjust the polyethylene thickness in 1 cm increments. The statistical uncertainty of the calculated $H^*(10)$ value should be within 5%.
- ✓ Task 3: Please create a more realistic snowman geometry based on *lecture/exercise/snowman*. Assign realistic colors using *[mat name color]*. Then, estimate the proton energy that maximizes the deposited energy in the target region of the snowman geometry, adjusting the energy in 1 MeV increments.
- ✓ Task 4: Please evaluate the correlations among the responses of three detectors. Use *sample/UserDefined/usrtally/t-deposit-event* as a reference. Construct a geometry with three cylindrical liquid scintillators placed consecutively along the same axis; each detector should have a radius of 10 cm and a thickness of 3 cm. Irradiate the detectors with 100 MeV neutrons along the central axis. Set the number of histories to 1000. Output the deposition energy in each of the three detectors for every event and calculate the probability that only the middle detector has a deposition energy above the 1 MeV threshold.
- ✓ Task 5: Please calculate the number of DNA damage events using KURBUC. Estimate the number of DNA damage events per source particle (damage/source) for 3 MeV protons incident on a water sphere with a radius of 50 nm. Use *sample/UserDefined/usrtally/DNAdamage* as a reference for the calculation method. Keep the number of histories unchanged from the sample input.

These tasks are hereafter referred to as cosmic-ray response, shielding optimization, melting-snowman exercise, detector coincidence, and DNA damage estimation, or simply as Tasks 1–5. Task 1 was designed to represent a typical PHITS simulation workflow, in which the geometry, source, and tally are appropriately defined, PHITS is executed, and the calculation results are obtained. Tasks 2 and 3 were optimization problems, in which PHITS simulations were performed repeatedly under different conditions to search for the optimal setting. Tasks 4 and 5 could not be handled by the standard tallies alone and required compilation of PHITS with user-defined tally functions. In particular, Task 5 was a highly advanced task that would conventionally have been feasible only for advanced users, because it required compilation with the specialized PHITS-KURBUC library [15] and compilation of a post-processing program to estimate DNA damage from track-structure analysis results obtained by PHITS [16].

Each of the five prompts was executed ten times independently, using different operating systems (Windows, macOS, and Ubuntu), AI agents (Codex, Claude Code, and GitHub Copilot), and LLMs (GPT-5.5, GPT-5.6, Opus 4.8, Opus 5, and Sonnet 5). For each completed task, the total computational time was recorded, and the AI processing time was then obtained by subtracting the execution time of PHITS and related programs from it. The computing environments and the computational times of all ten runs are summarized in Appendix II.

## 3. Results and Discussion

### *3.1. Generated PHITS knowledge base*

Table 1 summarizes the file size, word count, and character count of the text files included in the PHITS knowledge base stored in the *workbench/AI/knowledge_base* directory and uploaded to the Gemini Notebook page named AI Phi-chan. The knowledge base covers the PHITS manual in English and Japanese, the lecture materials, the sample input files and user-defined programs, the recommended-setting input files, instructions for using the supporting tools, and the developer-curated crucial notices. It also includes the manuals of the supporting tools, namely ANGEL, a graph-drawing program for PHITS, DCHAIN-PHITS [17], an inventory calculation code, and PHIG-3D [11]. These manuals are distributed in HTML format and were textualized in the same manner as the PHITS manual, and each of them is provided in both English and Japanese. In Gemini Notebook, the maximum length of a single source file is limited to 500,000 words; therefore, the lecture materials were divided into two files. The word counts were obtained by counting whitespace-delimited tokens and are therefore not directly comparable between languages, because Japanese does not separate words by spaces. The Japanese files consequently show smaller word counts than their English counterparts even though the content covered is almost identical. It should be noted that these data were generated based on PHITS version 3.37; therefore, the file sizes, word counts, character counts, and related values will be updated whenever a new version of PHITS is released.

### *3.2. Results of demonstration tasks*

Table 2 summarizes the answers obtained in the ten runs. Apart from the values marked with an asterisk, whose origins are examined below and in Appendix II, the agents carried out the intended calculations correctly, and the remaining differences reflect quantities that the prompts left unspecified. In Task 1, the number of histories was not prescribed and only statistical uncertainty below 10 % was required, so the answers scatter within that margin. In Task 3, the geometry itself was left to the agent, so the freedom is much greater, and the answers range from 122 to 380 MeV. Tasks 2, 4, and 5 were specified more tightly, and the unmarked runs gave essentially the same answer irrespective of the AI agent, language model, and operating system used.

The exceptionally low values obtained in Runs 9 and 10 for Task 1 originate from the definitions of the source and the geometry. In these runs, the isotropic spherical source was made to coincide exactly with a surface used in the cell definition, namely the outer surface of the aluminum shell. Under such conditions, PHITS may misidentify the source region. This behavior, together with the recommendation to shift the source slightly away from such surfaces, is described in the FAQ of the PHITS manual, which is part of the resources available to the agents; however, neither run found this description. Errors involving more than one input section, [source] and [surface] in this case, can easily be overlooked in a manual organized section by section. Cautions against such errors should therefore be stated in the crucial notices, which the agent reads before generating the input file.

In Task 2, seven runs answered 29 cm directly, and Run 9 answered 30 cm as the outer radius of a geometry containing a 1 cm vacuum cavity around the source, so that its polyethylene thickness is also 29 cm. Two runs answered 26 cm, in both cases because the secondary photons from $^{1}H(n,\gamma)^{2}H$ capture in the polyethylene were left out of the scored dose: Run 1 scored the neutron dose alone, while Run 3 neglected the photon transport. These cases have a common origin: where the prompt does not completely fix the quantity to be evaluated or the geometry to be built, the agent makes the decision itself, and the result may differ from what the user intended. None of these choices can be detected from the execution itself, so the user must read the input files and the summary files produced by the agents; this is an integral part of the workflow rather than an optional check.

The scatter in Task 3 is not a failure of the workflow: the prompt asked for a more realistic snowman geometry without prescribing it, so each run built a different body and hence a different depth of the target region. Within each run, the answer remained physically consistent, the optimum energy always corresponding to a proton range that reaches the distal edge of that run's own target. As an example, Figure 2 shows three-dimensional visualizations of the original geometry included in the snowman exercise and of the geometries generated in Runs 1 and 4, together with the particle flux or absorbed-dose distributions obtained at the optimal proton energy; Run 1 used a snowball density of 1 g/cm$^3$, as in the original exercise, whereas Run 4 adopted a more realistic value of 0.3 g/cm$^3$. An ambiguous instruction thus propagates directly into the answer even though every individual calculation is sound, and the spread in Table 2 reflects the freedom left by the prompt rather than the reliability of the agents.

Tasks 4 and 5 required PHITS itself to be rebuilt, and this succeeded in all ten runs: the agent installed a user-defined tally, compiled PHITS with it, and ran the resulting executable, and in Task 5 it additionally relinked the specialized PHITS-KURBUC library [15] and compiled the post-processing program that estimates DNA damage from the track-structure output [16]. With one exception, all the marked values in these two tasks originate in parts of the calculation that the prompts did not fix and that the agents therefore decided for themselves, such as source and geometry settings and the definition of the quantity to be reported. The exception is Run 8 in Task 5, where a build-dependent difference in the last significant digit of the event records changed the damage count by one, although its input file is identical to that of the other runs. More details, together with the computational times of the ten runs, are discussed in Appendix II.

None of these differences could have been detected from the criteria applied by the agents themselves: every run completed normally, reported zero lost particles and zero unrecovered geometry errors, and satisfied the statistical-uncertainty requirement wherever the prompt specified one. The differences were identified only by comparing the runs with one another and by re-examining the files generated by the agents. Successful execution and verification of statistical uncertainty alone therefore do not establish the physical validity of a simulation, and either independent repetition or human inspection of the summary files remains necessary.

### *3.3. Lessons learned for code users*

The prompts used in the demonstration tasks were not written perfectly from the beginning. Rather, they were refined through several trials in which unclear instructions caused the AI agents to perform simulations that were different from what we had intended. Based on these unsuccessful attempts, the following information should be included when asking an AI agent to perform Monte Carlo particle transport simulations:

- ✓ the purpose of the calculation;
- ✓ the name of the sample directory or input file to be used as a reference;
- ✓ the components to be modified and those to be kept fixed, such as the geometry, source, tally, and calculation parameters;
- ✓ the evaluation and optimization criteria, including the physical quantities to be evaluated and the required level of statistical uncertainty.

The third item is particularly important, because ambiguous instructions may cause AI agents to modify unintended parts of the input file. Users should therefore specify the conditions in as much detail as possible, for example whether only the size of a particular region should be modified, or whether only the source energy should be changed while keeping the spatial and angular distributions fixed.

It is efficient to provide all these items in a single prompt, but they can also be provided step by step if the user is unsure. For example, if the user does not know which sample should be used as a reference, the user can first describe the purpose of the calculation and ask the AI agent to suggest an appropriate sample. In many cases, the agent can identify a suitable reference input file. For optimization tasks that require repeated PHITS runs, it is also useful to first ask the agent to generate an input file for a specific condition and have it checked by a human before starting the repeated calculations. This procedure can reduce the risk of wasting computational time by repeatedly executing incorrect simulations. As an illustrative example of such step-by-step use, Appendix III presents a detailed interaction between a user and an AI agent in which a PHITS calculation for passive-beam proton therapy using an ICRP adult reference voxel phantom [18] was progressively developed, from the selection of an appropriate recommended input to beam optimization and organ-dose assessment.

Even if a calculation is completed successfully, as previously mentioned, human review, confirmation of statistical uncertainties, and assessment of physical validity are essential. A second AI agent, run either in a separate session or as a sub-agent, is also useful for checking the results. It is also important to save the PHITS execution logs, input files, and output files to ensure reproducibility. When an AI agent makes a mistake, the user should ask the agent to analyze the cause of the mistake and accumulate preventive cautions for future tasks. In the present workflow, this role is served by the *user/crucial_notice.txt* file.

### *3.4. Lessons learned for code developers*

Even if a user provides an appropriate prompt, the expected result cannot be obtained if the AI agent does not correctly understand the syntax and execution procedures of the particle transport code. In fact, during the demonstration tasks, the agents made several mistakes before the final prompts and reference policies were refined. Examples included using # as a comment character in the *[surface]* and *[cell]* sections, where it is not allowed, and proceeding to a particle transport calculation without checking a newly generated geometry and thereby causing lost-particle errors. Another example was running multiple PHITS input files in the same directory simultaneously, which resulted in file-conflict errors.

Based on these experiences, we consider the following points important for developers who wish to make particle transport codes more AI-friendly.

- ✓ Manuals and lecture materials should be provided not only in image-based or PDF-layout formats but also as text-based resources. Although AI systems can handle figures and images, input syntax, commands, keywords, file paths, and error messages are more reliably interpreted when they are provided as explicit text.
- ✓ A sufficient number of sample input files should be provided, together with descriptions of their purpose, assumptions, physics models, tallies, output files, modifiable parts, recommended usage, and cautions. For AI agents, high-quality samples serve as starting points for input generation. In particle transport simulations, it is generally safer to modify validated samples with minimal changes than to generate input files from scratch.
- ✓ Execution policies for AI agents should be clearly defined in advance. Working directories, file names, collision avoidance during repeated or parallel execution, execution commands, error-checking procedures, statistical-uncertainty checks, and geometry checks should be specified to prevent unintended agent behavior and accidental damage to files. These policies should also include typical mistakes to be checked and usages to be avoided, such as code-specific syntax restrictions, geometry-check requirements, file-conflict risks, and errors that arise from a combination of individually valid settings in different input sections.

- ✓ Execution paths controllable from a command-line interface or scripts should be provided in addition to graphical user interfaces. AI agents are well suited to file editing, command execution, log inspection, and script execution, but are less suited to manual GUI operations. Tools that can be executed from the command line, such as PHITS itself, ANGEL, DCHAIN-PHITS, and RT-PHITS, are therefore compatible with AI-agent workflows. For visualization tools such as PHIG-3D, a future task is to enable AI agents to control display settings through scripts or command-line options, so that image generation and geometry checking can be automated.

### *3.5. Limitations*

The present demonstration has two main limitations. First, the ten runs were carried out using the operating systems, AI agents, and LLMs available to the authors and were not designed as a systematic comparison. They therefore demonstrate only that the workflow can operate as intended in each of these environments, not how reliably an AI agent produces a correct answer. A more systematic evaluation would require repeating the same prompts on the same computer using a variety of AI agents and quantitatively assessing the variability in their results.

Second, the workflow was demonstrated using only five prompts, which covered a limited portion of PHITS functionality. None of the tasks involved supporting software distributed with PHITS, such as DCHAIN-PHITS for activation analysis or RT-PHITS for radiotherapy applications. In addition, only English-language prompts were tested. The performance of the workflow in other languages, particularly in languages for which no PHITS manual is available, therefore remains to be evaluated.

## 4. Conclusions

In this study, we presented a code-side implementation strategy for applying existing AI assistants and AI agents to PHITS. Two complementary sets of AI-ready resources were introduced in the *workbench* directory: a bundled knowledge base for RAG-based AI assistants and a compact agent reference for direct use by AI agents. Together with agent instruction files, PHITS-specific reference policies, execution rules, task definitions, and setup tools, these resources enabled conversational PHITS support and AI-agent-driven execution and analysis workflows. A key feature of this approach is that the resources and operational rules required for AI-assisted workflows are bundled with the PHITS package, rather than implemented in a dedicated PHITS-specific AI application. This allows existing general-purpose AI tools to support usage-related questions, input-file preparation, simulation execution, result verification, and analysis. The primary contribution of this study is therefore not the demonstration that AI agents can operate PHITS, but the practical methodology it establishes for preparing the code-side AI-ready resources and operational rules required for particle transport simulation codes.

Although this study demonstrated a code-side implementation strategy for applying existing AI tools to PHITS, further work is needed to generalize and optimize this approach for broader particle transport simulation workflows. As part of this effort, future studies should establish design principles and evaluation metrics for AI-ready knowledge bases, agent references, sample files, and execution policies that enable AI agents to perform particle transport simulations with fewer errors. Visualization and other supporting tools should also be made accessible to AI agents through scripts or command-line interfaces, thereby enabling automated geometry checking, result visualization, and related operations. The practical effectiveness of such workflows should then be evaluated quantitatively using different AI agents, in terms of task success rates, error-correction capability, human working time, and applicability to different simulation fields.

If these future studies establish effective AI-assisted workflows for many particle transport codes, the practical boundaries between codes may gradually become less important for users. In such a future, users may be able to execute multiple particle transport codes, compare their results, and analyze the differences simply by submitting a single prompt to a single AI agent. Furthermore, the AI agent may be able to select the most appropriate code and computational resources according to the purpose of the calculation and the available hardware, for example by choosing a GPU-accelerated Monte Carlo code when computational speed is a primary requirement.

**Declaration of generative AI use**

During the preparation of this manuscript, the authors used ChatGPT (OpenAI) and Claude Code (Anthropic) to assist with English editing, wording suggestions, and improvement of clarity. The authors reviewed and edited all AI-assisted outputs and take full responsibility for the content of the manuscript.

**Disclosure statement**

No potential conflict of interest was reported by the authors.

**Data availability statement**

The data that support the findings of this study are available from the corresponding author upon reasonable request.

**Figure captions**

Figure 1. Overview of the PHITS package with AI-assisted workflow components introduced in this study.

Figure 2. PHIG-3D visualizations of the original geometry included in the snowman exercise and the geometries generated by Runs 1 and 4 (upper panels), together with the particle flux or absorbed-dose distributions calculated at the optimized proton energy for each geometry (lower panels).

Table 1. File size, word count, and character count of the text files included in the PHITS knowledge base.

| File name | Description | File size | #Words | #Characters |
|---|---|---|---|---|
| angel-en_forAI.txt | Textualized ANGEL manual in English | 310 kB | 28.9k | 317k |
| angel-jp_forAI.txt | Textualized ANGEL manual in Japanese | 310 kB | 15.3k | 239k |
| crucial_notice_forAI.txt | Crucial notice for AI | 3.67 kB | 0.607k | 3.76k |
| dchain-en_forAI.txt | Textualized DCHAIN-PHITS manual in English | 360 kB | 42.4k | 365k |
| dchain-jp_forAI.txt | Textualized DCHAIN-PHITS manual in Japanese | 357 kB | 25.2k | 282k |
| lecture_forAI_part01.txt | Textualized lecture materials (Part I) | 3.11 MB | 431k | 2.99M |
| lecture_forAI_part02.txt | Textualized lecture materials (Part II) | 1.30 MB | 197k | 1.31M |
| manualE_forAI.txt | Textualized PHITS manual in English | 1.02 MB | 161k | 1.07M |
| manualJ_forAI.txt | Textualized PHITS manual in Japanese | 1.20 MB | 68.0k | 713k |
| phig3d-en_forAI.txt | Textualized PHIG-3D manual in English | 30.7 kB | 4.53k | 31.3k |
| phig3d-ja_forAI.txt | Textualized PHIG-3D manual in Japanese | 36.4 kB | 0.848k | 17.6k |
| recommendation_forAI.txt | Recommended-setting input files | 209 kB | 22.8k | 179k |
| sample_forAI.txt | Sample input files and user-defined programs | 2.43 MB | 264k | 2.49M |
| utility_forAI.txt | Instructions for using supporting tools | 1.26 MB | 121k | 1.30M |
| | Total | 11.9 MB | 1.38M | 11.3M |

Table 2. Answers obtained for the five demonstration tasks in the ten independent runs. Tasks 1–5 are the cosmic-ray response, shielding optimization, melting-snowman exercise, detector coincidence, and DNA damage estimation, respectively. The computing environment and the computational times of each run are given in Appendix II.

| Run | AI agent (LLM) | Task 1 (counts/s) | Task 2 (cm) | Task 3 (MeV) | Task 4 (events/source) | Task 5 (damage/source) |
|---|---|---|---|---|---|---|
| 1 | Codex (GPT-5.5) | 23.2 | 26* | 158 | 0.026 | 0.60 |
| 2 | Codex (GPT-5.5) | 23.9 | 29 | 122 | 0.026 | 0.60 |
| 3 | Codex (GPT-5.6) | 23.5 | 26* | 195 | 0.026 | 0.60 |
| 4 | Claude Code (Opus 4.8) | 24.3 | 29 | 131 | 0.026 | 0.60 |
| 5 | Claude Code (Opus 5) | 23.8 | 29 | 242 | 0.026 | 0.60 |
| 6 | Claude Code (Sonnet 5) | 26.2 | 29 | 248 | 0.026 | 0.64* |
| 7 | Claude Code (Opus 5) | 23.6 | 29 | 180 | 0.022* | 0.62* |
| 8 | Claude Code (Sonnet 5) | 25.8 | 29 | 320 | 0.026 | 0.56* |
| 9 | Claude Code (Sonnet 5) | 7.98* | 30* | 380 | 0.026 | 0.60 |
| 10 | GitHub Copilot (Sonnet 5) | 8.56* | 29 | 132 | 0.026 | 0.36* |

* The reasons why these values differ from the others are identified in Section 3.2 and Appendix II.

# Appendix I. Strategies for Preparing the AI-Ready Knowledge Base and Agent Reference

This appendix describes the strategy used in this study to prepare AI-ready reference resources from the PHITS package, which may also be applicable to other particle transport codes. The purpose was to reorganize the existing manuals, lecture materials, sample inputs, recommended calculations, utility programs, and developer-curated cautions into formats that can be efficiently searched and used by both retrieval-augmented generation (RAG) systems and AI agents. The original PHITS files were retained in their original locations; the AI-ready resources were generated as derived files by Python scripts.

Two complementary outputs were prepared: a knowledge_base directory and an agent_reference directory. The knowledge_base directory was intended for RAG-based AI assistants. It contains multiple merged text bundles, each corresponding to a major source category, such as manuals, lectures, utilities, sample inputs, and recommended calculations. The bundles include file indexes, extracted document text, source paths, metadata, and, where appropriate, the contents of related input and supplemental files. Large collections of lecture materials were divided into multiple files to keep individual bundles manageable.

The conversion scripts first scanned the relevant PHITS directories and prepared indexes of the available materials. Each entry records the source path and, where possible, a short description of its purpose. Text was extracted from DOCX, PPTX, and PDF files. For PPTX files, both slide text and speaker notes were collected when available. The manual sources were converted while retaining their section hierarchy, including linked or included files. These procedures made information originally distributed across different document formats accessible through text search.

Additional metadata were prepared to improve retrieval. For individual input files, the intended purpose was inferred from title sections, explanatory comments, folder names, or file names. Keywords were generated from informative section names while excluding generic structural sections. The resulting metadata help AI systems identify examples related to specific calculation purposes, particle types, geometries, tallies, or output quantities. Related instructional documents and small supplemental text or program files were included when they were useful for interpreting an example. In contrast, binary files, images, calculation outputs, and excessively large auxiliary files were excluded or represented only by their paths.

The agent_reference directory was prepared for direct use by AI agents operating in the PHITS workspace. It contains a compact top-level index and smaller Markdown files divided by category, source folder, or utility. These Markdown files provide extracted documentation, searchable headings, keywords, purpose descriptions, and paths to the original materials. However, input files, output files, source programs, and other plain-text files were generally not duplicated in this directory. Because an AI agent can directly read or copy the original files when necessary, duplication would unnecessarily increase token consumption. This lightweight structure supports targeted searches and selective reading during AI-agent-driven PHITS workflows.

# Appendix II. Details of Five Demonstration Tasks

This appendix summarizes the computing environments, the computational times, and the detailed causes of the differences among the ten runs of the five demonstration tasks discussed in Section 3.2. Table A2.1 lists the AI agent, the LLM, and the computer used in each run. Three runs used Codex, six used Claude Code, and one used GitHub Copilot; six were performed on Windows 11, three on Ubuntu, and one on macOS. Runs 1 and 4 were performed on the same computer. Each run was conducted independently of the other runs. Within each run, each task was conducted in a separate agent session, except for Run 3, in which all five tasks were carried out consecutively within a single agent session.

Table A2.2 lists the computational time of each task. The total computational time is the wall-clock time needed to complete the task, excluding time spent waiting for user approval. The AI processing time given in parentheses was obtained by subtracting the execution time of PHITS and the related programs from the total computational time. These times are not a measure of the performance of the agent. They depend on the speed of the computer and on the number of threads used, and above all on decisions that the prompts left to the agent, namely the number of histories in each simulation and the number of simulations repeated. The extreme case is Task 2 of Run 10, which took 9422 s because the agent followed a coarse scan of the shield radius with production runs of $1.5 \times 10^6$ and $3 \times 10^6$ histories, whereas Run 6 completed the same task in 1110 s by running PHITS with ten OpenMP threads. In Tasks 4 and 5, the transport calculation itself accounts for only up to about two minutes of execution time, so almost all of the program execution time is the time needed to rebuild PHITS with the user-defined tally and, in Task 5, to relink the KURBUC library and compile the DNA-damage post-processor.

The remaining values marked with an asterisk in Table 2, whose causes are not described in Section 3.2, are listed below together with the setting that each run interpreted differently. In Task 4, Run 7 placed the three scintillators with 1 cm vacuum gaps between them instead of in contact. In Task 5, Run 6 lowered the cut-off energy of the electron track structure from 7 eV to 1 eV, referring to a sample input file contained in the lecture materials for the track-structure mode instead of the DNA-damage sample that the prompt gave as the reference. Run 7 obtained its answer from the numbers of damage events of each type that the DNA-damage analysis program writes as an intermediate result, rather than from the total damage count used by the other runs, so that its definition of a DNA damage event differs from theirs. Run 10 widened the source from a pencil beam on the axis to a disc covering the whole cross-section of the target.

**Table A2.1.** Computing environment, AI agent, and LLM used for each of the ten runs.

| Run | AI agent (LLM) | OS | CPU | RAM |
|---|---|---|---|---|
| 1 | Codex (GPT-5.5) | Windows 11 | Core i7-9700K @ 3.60 GHz | 32 GB |
| 2 | Codex (GPT-5.5) | Windows 11 | Ryzen 5 7430U | 16 GB |
| 3 | Codex (GPT-5.6) | Ubuntu 24.04 | Xeon Gold 6226R @ 2.90 GHz | 32 GB |
| 4 | Claude Code (Opus 4.8) | Windows 11 | Core i7-9700K @ 3.60 GHz | 32 GB |
| 5 | Claude Code (Opus 5) | Windows 11 | Core i7-9700 @ 3.00 GHz | 16 GB |
| 6 | Claude Code (Sonnet 5) | Ubuntu 24.04 | Xeon E5-2698 v3 @ 2.30 GHz | 64 GB |
| 7 | Claude Code (Opus 5) | Ubuntu 26.04 | Ryzen 9 9950X | 256 GB |
| 8 | Claude Code (Sonnet 5) | Windows 11 | Core i5-14400 @ 2.50 GHz | 16 GB |
| 9 | Claude Code (Sonnet 5) | Windows 11 | Core i7-1065G7 @ 1.30 GHz | 32 GB |
| 10 | GitHub Copilot (Sonnet 5) | macOS 26.6 | Apple M4 Pro | 48 GB |

**Table A2.2.** Computational time of each task in each run. The upper value is the total computational time in seconds and the value in parentheses is the AI processing time.

| Run | Task 1 | Task 2 | Task 3 | Task 4 | Task 5 |
|---|---|---|---|---|---|
| 1 | 387<br>(177) | 2248<br>(212) | 757<br>(218) | 488<br>(461) | 622<br>(380) |
| 2 | 801<br>(205) | 1114<br>(555) | 1524<br>(510) | 413<br>(371) | 447<br>(374) |
| 3 | 602<br>(504) | 540<br>(231) | 421<br>(239) | 383<br>(380) | 299<br>(293) |
| 4 | 634<br>(400) | 1657<br>(403) | 1307<br>(830) | 354<br>(315) | 1194<br>(912) |
| 5 | 1088<br>(701) | 2991<br>(852) | 2557<br>(919) | 423<br>(394) | 828<br>(634) |
| 6 | 649<br>(634) | 1110<br>(967) | 2666<br>(1703) | 529<br>(516) | 523<br>(497) |
| 7 | 1335<br>(661) | 2997<br>(661) | 1385<br>(1221) | 580<br>(534) | 1066<br>(970) |
| 8 | 828<br>(814) | 720<br>(586) | 1490<br>(1052) | 661<br>(467) | 425<br>(394) |
| 9 | 476<br>(448) | 995<br>(705) | 2431<br>(1319) | 78<br>(52) | 462<br>(445) |
| 10 | 465<br>(460) | 9422<br>(1023) | 1260<br>(1074) | 363<br>(329) | 540<br>(525) |

# Appendix III. Illustrative User–AI Agent Interaction for PHITS-Based Proton Therapy Dose Assessment

This appendix illustrates a step-by-step interaction between a user and an AI agent for developing a PHITS-based dose assessment of passive-beam proton therapy. Codex (GPT-5.5) was used as the sole AI agent in this example. The elapsed times shown in parentheses indicate the actual wall-clock time required for each response, including both the time spent by Codex and the time required to execute PHITS and related programs.

It should be noted that the PHITS package already contains sample input files for a ridge filter and ICRP adult reference voxel phantoms. Therefore, Codex did not construct these components entirely from scratch, but instead adapted and combined existing PHITS resources according to the user’s instructions. Figure A3.1 shows the final geometry and particle tracks for 100 source particles rendered using PHIG-3D. Because PHIG-3D cannot yet be controlled directly by the AI agent, this visualization was generated manually by the authors after completion of the agent-assisted workflow.

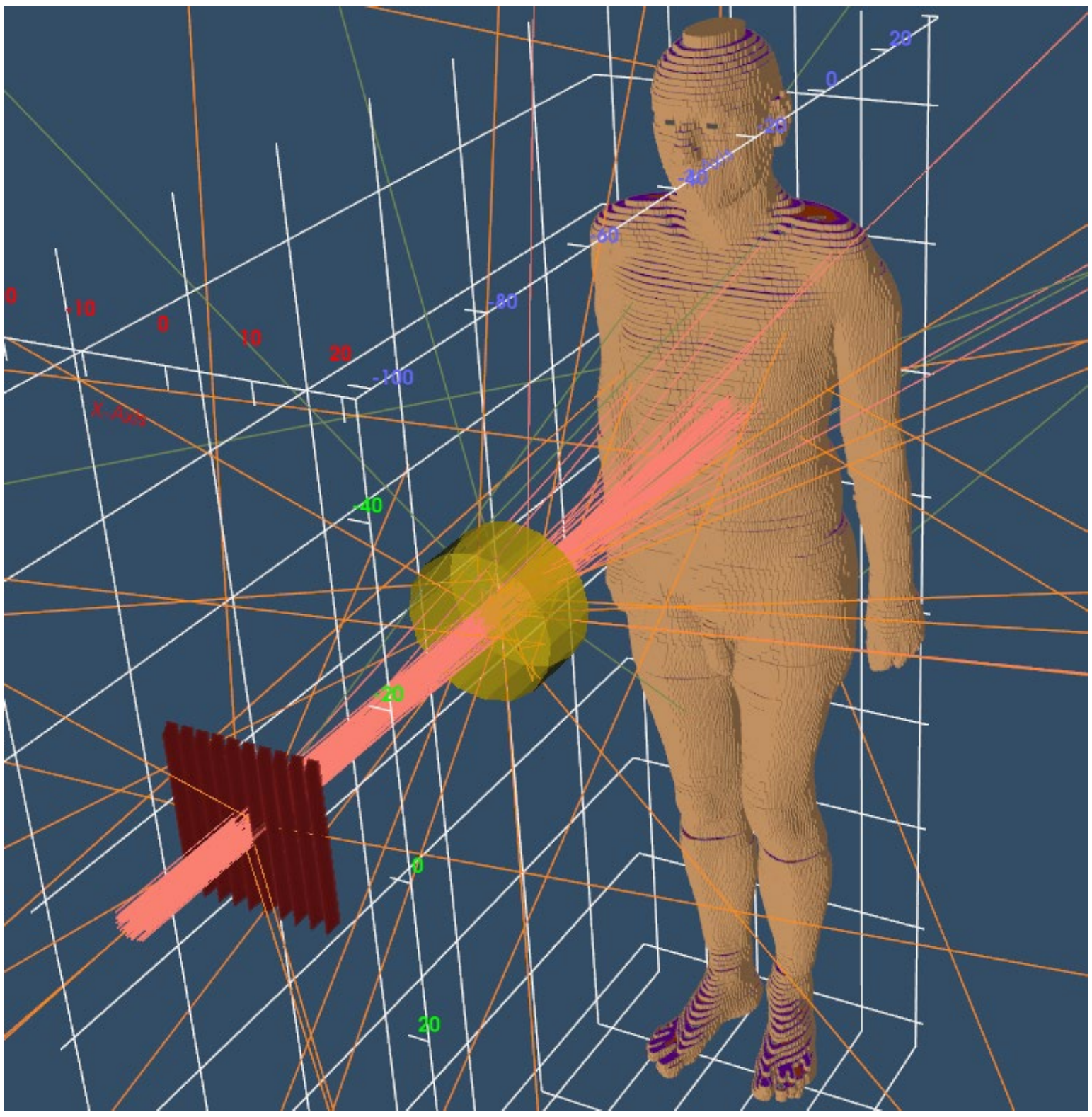

Figure A3.1. Final geometry and particle tracks for 100 source particles rendered using PHIG-3D.

**User-AI Agent Interaction Log**

User:
I would like to perform a calculation related to the medical physics of proton therapy. Please tell me the appropriate recommendation settings.

AI Agent (47 s):
I checked the local PHITS references and recommended settings, and identified the particle therapy recommendation input as an appropriate starting point for proton therapy depth-dose calculations.

User:
Copy the recommendation input file to user/protontherapy/water_phantom.inp, and calculate the depth-dose distribution for 200 MeV protons incident on a cylindrical water phantom. Set the cylinder radius to 10 cm and the depth to 30 cm.

AI Agent (4 m 14 s):
I copied the recommended input file to user/protontherapy/water_phantom.inp and modified it for 200 MeV protons incident on a cylindrical water phantom. The phantom was set to a radius of 10 cm and a depth of 30 cm, with a tally configured to calculate the depth-dose distribution.

User:
Set the beam radius to 2 cm, and adjust the tally region to match this range.

AI Agent (2 m 41 s):
I changed the beam radius to 2 cm and adjusted the tally region to match the beam range.

User:
Place a ridge filter 10 cm upstream of the irradiation surface, and adjust the ridge filter geometry so that the SOBP width becomes 5 cm. Adjust the number of histories so that the statistical uncertainty in the SOBP region is 2% or less. The absorbed dose variation within the SOBP region should be kept within 5%.

AI Agent (29 m 53 s):
I placed a ridge filter 10 cm upstream of the irradiation surface and adjusted the ridge-filter geometry so that the SOBP width became approximately 5 cm. I also increased the number of histories so that the statistical uncertainty in the SOBP region was 2% or less, and checked that the absorbed dose variation within the SOBP region was within 5%.

User:
Move the ridge filter to a position 1 m upstream of the irradiation surface, and place a PMMA range shifter with a radius of 10 cm and a thickness of 1 cm halfway between the ridge filter and the irradiation surface. Then, check whether the SOBP satisfies the above conditions. If it does not, adjust the ridge filter geometry again.

AI Agent (24 m 30 s):
I moved the ridge filter to 1 m upstream of the irradiation surface and placed a PMMA range shifter with a radius of 10 cm and a thickness of 1 cm halfway between the ridge filter and the irradiation surface. I then checked whether the SOBP satisfied the previous conditions and adjusted the ridge-filter geometry as needed.

User:
Without changing the source conditions at all, replace the water phantom with the ICRP adult male voxel phantom. Assuming stomach cancer, adjust the position and orientation of the phantom so that the proton SOBP beam is delivered from the anterior side of the body toward the stomach.

AI Agent (33 m 11 s):
Done. I replaced the water phantom in C:/phits/user/protontherapy/water_phantom.inp with the ICRP adult male voxel phantom and copied the text lattice to C:/phits/user/protontherapy/male-lattice.inp.
I preserved the source conditions unchanged: 200 MeV proton, 2 cm radius, z0/z1 = -120 cm, dir = 1.

The phantom was transformed so that the fixed +z beam enters from the anterior side and points toward the stomach. I added stomach and organ dose tallies.
Verification:
- The icntl=8 geometry check completed, with no *_geo.out files produced.
- The full PHITS run completed: PHITS simulation with icntl = 0 is properly finished.
- The stomach dose uncertainty was well below 2%:
  - Stomach wall region 72: 9.9074E+00, r.err 0.0028
  - Stomach contents region 73: 1.9990E+01, r.err 0.0028
One caveat: the completed run reported recoverable geometry warnings and one lost particle in the ridge-filter region, not in the voxel phantom. PHITS still finished properly with Number of unrecovered errors = 0.

User:
Adjust only the thickness of the range shifter so that the dose to the stomach contents is maximized.

AI Agent (3 h 31 m 16 s):
Done. I adjusted only the PMMA range-shifter thickness in C:/phits/user/protontherapy/water_phantom.inp.
Final setting:
- Range shifter thickness: 9.25 cm
- Surfaces: z = -54.625 cm to z = -45.375 cm
- Center remains at z = -50 cm
I swept the thickness and found that the stomach-contents dose peaked around 9.0-9.25 cm; 9.25 cm was the highest value in the refined sweep. The production run with 100,000 histories finished properly.
Final stomach contents dose:
- Region 73: 2.4144E+01 MeV/source
- Relative error: 0.0036
For comparison, the previous 1 cm range shifter gave 1.9990E+01 MeV/source, so this increased the stomach-contents dose by about 21%.
The source conditions were unchanged. There was still the pre-existing recoverable ridge-filter geometry warning with one lost particle and zero unrecovered errors.

User:

Assuming a beam current of 1 nA, calculate the irradiation time required to deliver an absorbed dose of 10 Gy to the stomach contents under this irradiation condition.

Also, report the absorbed doses to the surrounding organs for this irradiation time.

AI Agent (1 m 55 s):

Using the final 9.25 cm PMMA range-shifter condition:

- Stomach contents energy deposition: 24.144 MeV/source
- Stomach contents volume: 240.358 cm3
- Density: 1.04 g/cm3
- Mass: 0.24997 kg
- Beam current: 1 nA = 6.2415E+09 protons/s

Dose rate to stomach contents:

0.09659 Gy/s

Therefore, the irradiation time required to deliver 10 Gy to the stomach contents is:

103.5 s = 1.73 min

Approximate absorbed doses for the same irradiation time:

| Organ / tissue | Absorbed dose |
|---|---|
| Stomach contents | 10.0 Gy |
| Stomach wall | 7.25 Gy |
| Pancreas | 1.85 Gy |
| Small intestine | 0.727 Gy |
| Liver | 0.562 Gy |
| Spleen | 0.301 Gy |
| Large intestine | 0.217 Gy |
| Oesophagus | 0.210 Gy |
| Adipose tissue | 0.169 Gy |
| Kidneys | 0.0118 Gy |
| Urinary bladder | 0.00480 Gy |

These are mass-averaged organ/material doses scaled from male-organ-dose.out. The target timing used the dedicated stomach-dose.out tally for region 73.

User:

For this condition, I would like to visualize the particle tracks for 100 source particles using PHIG-3D. Please add [t-4dtrack] and run PHITS with a short history number.

AI Agent (6 m 4 s):

Done. I created and ran the PHIG-3D track case here:

C:/phits/user/protontherapy/phig3d_tracks/track_100.inp

Changes in that visualization input:

- maxcas = 100
- maxbch = 1
- Added [ T-4Dtrack ]
- Output file: 4Dtrack.out
- HistoryMax = 100
- verbose = 0

PHITS completed properly and produced:

C:/phits/user/protontherapy/phig3d_tracks/4Dtrack.out

Run check:

- PHITS simulation with icntl = 0 is properly finished
- Lost particles: 0
- Unrecovered geometry errors: 0
- No *_geo.out files were produced in the run folder.

I also launched PHIG-3D with track_100.inp. In PHIG-3D, use the Particle tracks tab and open 4Dtrack.out to display the 100-source-particle trajectories.